\documentclass[aps,pre,showpacs,amsmath,amsfonts,amssymb,superscriptaddress,preprint]{revtex4}

\usepackage[english]{babel}
\usepackage{graphicx}

\usepackage{amsthm}
\usepackage{amssymb}
\usepackage{SIunits}
\usepackage{amsmath,mathrsfs}
\usepackage{hhline}
\usepackage{fancyhdr}

\usepackage{appendix}

\everymath{\displaystyle}
\everymath{\displaystyle}

\newcommand{\affA}{Aix Marseille Universit\'{e}, Inserm, Institut de Neurosciences des Syst\`{e}mes, UMR$\_$S 1106, 13005, Marseille, France}
\newcommand{\affB}{Facult\'{e} de M\'{e}decine de la Timone, centre de R\'{e}sonance Magn\'{e}tique et Biologique et M\'{e}dicale (CRMBM, UMR CNRS-AMU 7339), Medical School of Marseille, 
Aix-Marseille Universit\'{e}, 13005, Marseille, France}
\newcommand{\affC}{Assistance Publique - H\^{o}pitaux de Marseille, H\^{o}pital de la Timone, P\^{o}le d'Imagerie, CHU, 13005, Marseille, France}
\newcommand{\affD}{Assistance Publique - H\^{o}pitaux de Marseille, H\^{o}pital de la Timone, Service de Neurophysiologie Clinique, CHU, 13005 Marseille, France}
\newcommand{\affE}{Inria Sophia Antipolis M\'{e}diterran\'{e}e Research Centre, MathNeuro Team, 2004 route des Lucioles-Bo\^{\i}te Postale 93 06902 Sophia Antipolis, Cedex, France}
\newcommand{\affG}{CNR - Consiglio Nazionale delle Ricerche - Istituto dei Sistemi Complessi, 50019, Sesto Fiorentino, Italy}

\begin{document}

\title{Controlling seizure propagation in large-scale brain networks}

\author{Simona Olmi}
\email{simona.olmi@inria.fr}
\affiliation{\affE}\affiliation{\affG}
\author{Spase Petkoski}
\affiliation{\affA}
\author{Maxime Guye}
\affiliation{\affB}\affiliation{\affC}
\author{Fabrice Bartolomei}
\affiliation{\affD}
\author{Viktor Jirsa}
\affiliation{\affA}

\begin{abstract}
Information transmission in the human brain is a fundamentally dynamic network process. In partial epilepsy, this process is perturbed and highly synchronous seizures originate in a 
local network, the so-called epileptogenic zone (EZ), before recruiting other close or distant brain regions. We studied patient-specific brain network models of 15 drug-resistant 
epilepsy patients with implanted stereotactic electroencephalography (SEEG) electrodes. Each personalized brain model was derived from structural data of magnetic resonance imaging 
(MRI) and diffusion tensor weighted imaging (DTI), comprising 88 nodes equipped with region specific neural mass models capable of demonstrating a range of epileptiform discharges. 
Each patient’s virtual brain was further personalized through the integration of the clinically hypothesized EZ. Subsequent simulations and connectivity modulations were performed 
and uncovered a finite repertoire of seizure propagation patterns. Across patients, we found that (i) patient-specific network connectivity is predictive for the subsequent seizure 
propagation pattern; (ii)seizure propagation is characterized by a systematic sequence of brain states; (iii) propagation can be controlled by an optimal intervention on the connectivity matrix; 
(iv) the degree of invasiveness can be significantly reduced via the here proposed seizure control as compared to traditional resective surgery. To stop seizures, neurosurgeons typically 
resect the EZ completely. We showed that stability analysis of the network dynamics using graph theoretical metrics estimates reliably the spatiotemporal properties of seizure propagation. 
This suggests novel less invasive paradigms of surgical interventions to treat and manage partial epilepsy.  
\end{abstract}
\date{\today}
\maketitle


Propagation of activity through a network is a non-stationary spatiotemporal process and is the most fundamental representation of information processing in the brain \cite{Kirst2016, Palmigiano2017}. 
In task conditions, as the behavioral dynamics unfolds, brain activity simultaneously evolves in a hierarchy of characteristic network activations. For instance, decision making tasks involve a chain of 
sequential subnetwork activations comprising the initial preparatory phase, the decision phase and the final decision execution phase \cite{Roelfsema, Alexa_Riehle}. Information processing in sensorimotor 
coordination \cite{Daffertshofer} and auditory, visual and linguistic tasks \cite{Barry_Horwitz} show robust propagation through well-timed activation chains of characteristic subnetworks. 

In the diseased and aging brain, the spatiotemporal organization of information processing is disrupted \cite{Sleimen-Malkoun2015, Battaglia2018}. Local and distributed alterations of 
connectivity and cell tissue properties influence the functional capacity of the brain in stroke, schizophrenia, multiple sclerosis and a range of neurodegenerative disorders 
\cite{Destexhe2016, Isomura2008, Wendling2002, Bernard2000, Regis2017}. 
From the perspective of spatiotemporal signal propagation and cognitive networks, epilepsy takes a key role, since seizure propagation is often accompanied by the evolution of a characteristic 
behavioral pattern, the semiology, which unfolds behaviorally as the neuroelectric activity is traced out in space and time. Epilepsy is a common disorder, affecting over 65 million people worldwide 
\cite{Ngugi2010}, frequently accompanied by cognitive and memory deficits \cite{Blake2000, Wang2011}. The investigation of electrographic spatiotemporal patterns (including interictal spikes, 
rhythmic bursts, wave propagation) relative to cognition networks dates back to 1939 \cite{Schwab1939} and is a crucial step in improving treatment and quality of life of patients with epilepsy. 

The influence of network topology and the anatomical organization of the epileptogenic process are particularly important in the context of seizure control and epilepsy surgery \cite{Bartolomei2013}. 
Bancaud and Talairach called the region of seizure onset the ``epileptogenic zone'' (EZ) \cite{Talairach1966}, 
which is not necessarily identical to the structural lesion and seizures have been reported to arise from structures distant from the lesion itself \cite{Talairach1966}. Often the clinical manifestations 
do not occur with seizure onset, but actually with the propagation of the seizure to other regions, which places further importance upon the understanding of how the spatiotemporal organization and spread 
of discharge activity arises. The concept that focal epilepsies are indeed not so local and involve large scale macroscale circuits is becoming increasingly more accepted in epileptology 
\cite{Spencer2002, Guye2008, Wendling2010, Laufs2012, Bartolomei2013, daSilva2013}. 

Practically, however, there are fundamental obstacles to the application of network control in the spatiotemporal organization and propagation of healthy and pathological activity. Antiepileptic drugs 
are most common and act globally by mostly targeting ion channels and inhibitory neurotransmission and synaptic receptors across the entire brain network. Yet approximately 30$\%$ of epilepsy patients 
suffer from drug-resistant epilepsy and continue to experience 
seizures despite appropriate antiepileptic drug treatment \cite{Kwan2011}. 
In recent years, cortical and subcortical neurostimulation is emerging as a promising treatment for drug-resistant epilepsy \cite{Fisher2012, Fisher2014}.
The remaining most common network control technique is resective surgery, in which the surgeon removes the epileptogenic zone (ascribed to temporal and extratemporal resections and hemispherectomy). 
The second, less common type of epilepsy surgery interrupts nerve pathways that allow seizures to propagate. This procedure, termed disconnection, is used for corpus callosotomy, hemispherotomy and 
multiple subpial transections \cite{disconnection}. 
The application of disconnection to other brain regions has been largely unsuccessful \cite{disconnection} and the causes of failure remain unknown. 
With the advent of novel minimally invasive surgery techniques, microsurgical disconnections become realistic to be performed in the human and present the least invasive restorative intervention for 
brain network control \cite{Gonzalez-Martinez2016}. These techniques include laser surgery via thermal ablation and Gamma knife radio surgery, in which converging narrow ionizing beams are focused 
to destroy cell tissue. Whether the goal of the network intervention is the suppression of seizure activity or alleviation of cognitive deficits due to comorbidities, conditio sine qua non for will 
always be a good understanding of the consequences of the network manipulation. 

To address this issue in models and patient brains, we here construct patient-specific connectome-based large-scale brain network models and identify the least invasive strategies to stop seizure propagation. 
Personalized connectomes are composed by the structural links derived from 
diffusion weighted tensor imaging (DTI) and tractography, and have been shown to predict individual functional connectivity better than generic approaches. Empirical studies are impossible to be 
performed systematically in the human and are limited to sparse sampling (empirical realizations) only. In silico modeling allows to complete the analysis with high-density sampling 
(all possible in-silico realizations) and identify the best network modulation for network control. We adapt a disconnection approach, aiming to identify the minimal ablation for a particular 
patient to stop the propagation of activity through the personalized network, let this be propagation due to stimulation or spontaneous seizures. Neuroelectric discharges are captured by gold 
standard data of epileptic patients recorded via intracranial electrodes. Brain areal dynamics is derived from the mean activity of neural populations via neural mass models able to capture the 
details of the autonomous slow evolution of interictal and ictal phases. Personalized brain network models are then derived for 15 patients. We systematically investigate 
all possible realizable disconnections of the epileptogenic zone and demonstrate that for all patients a partial targeted disconnection, accounting for the resection of one or a few pathways, 
is sufficient to stop seizure propagation in the brain, systematically reducing invasiveness and outperforming generic (non-personalized) procedures. 

\section{Results}

\begin{figure}
\begin{center}
\includegraphics[width=0.39\textwidth]{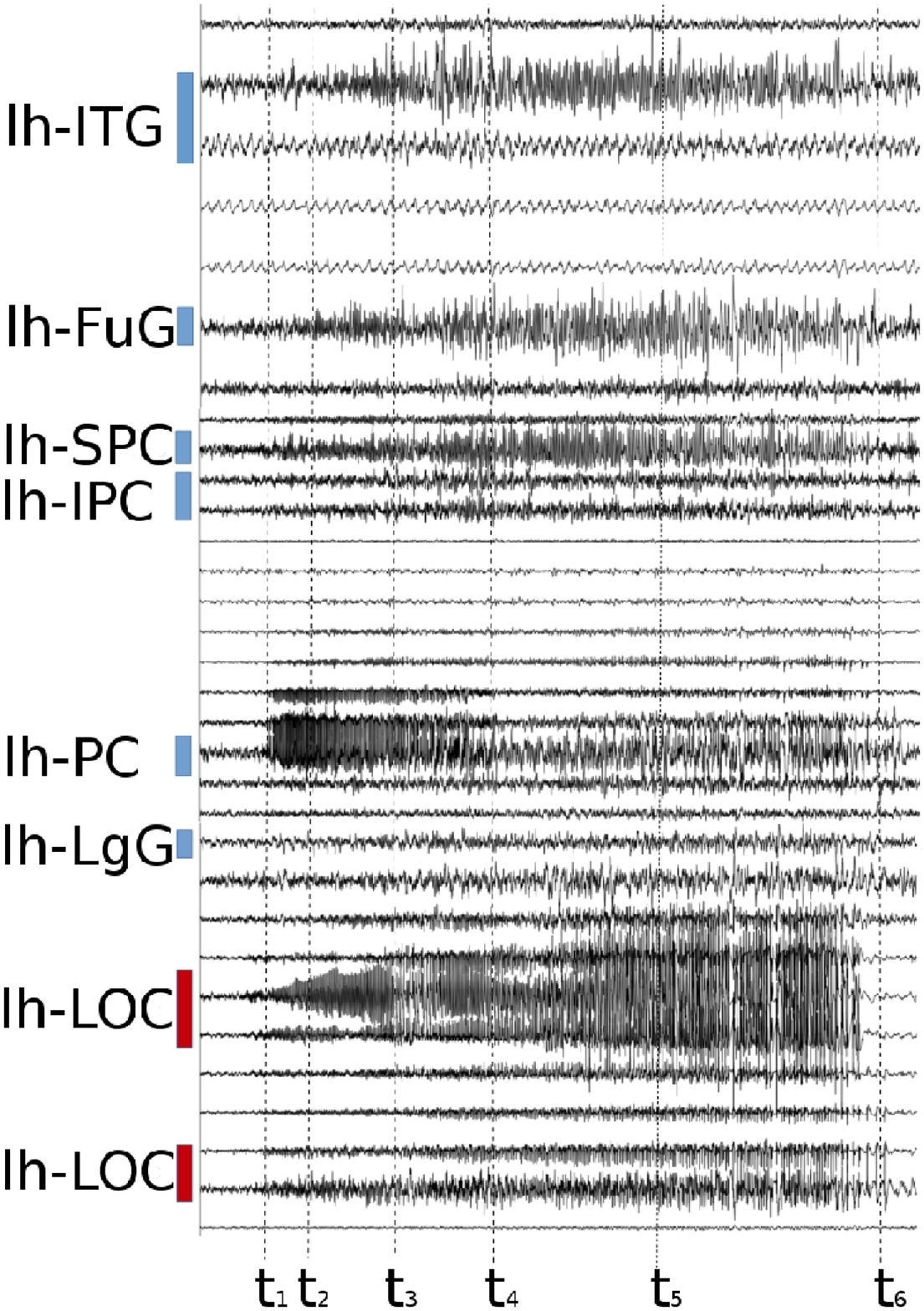}
\includegraphics[width=0.27\textwidth]{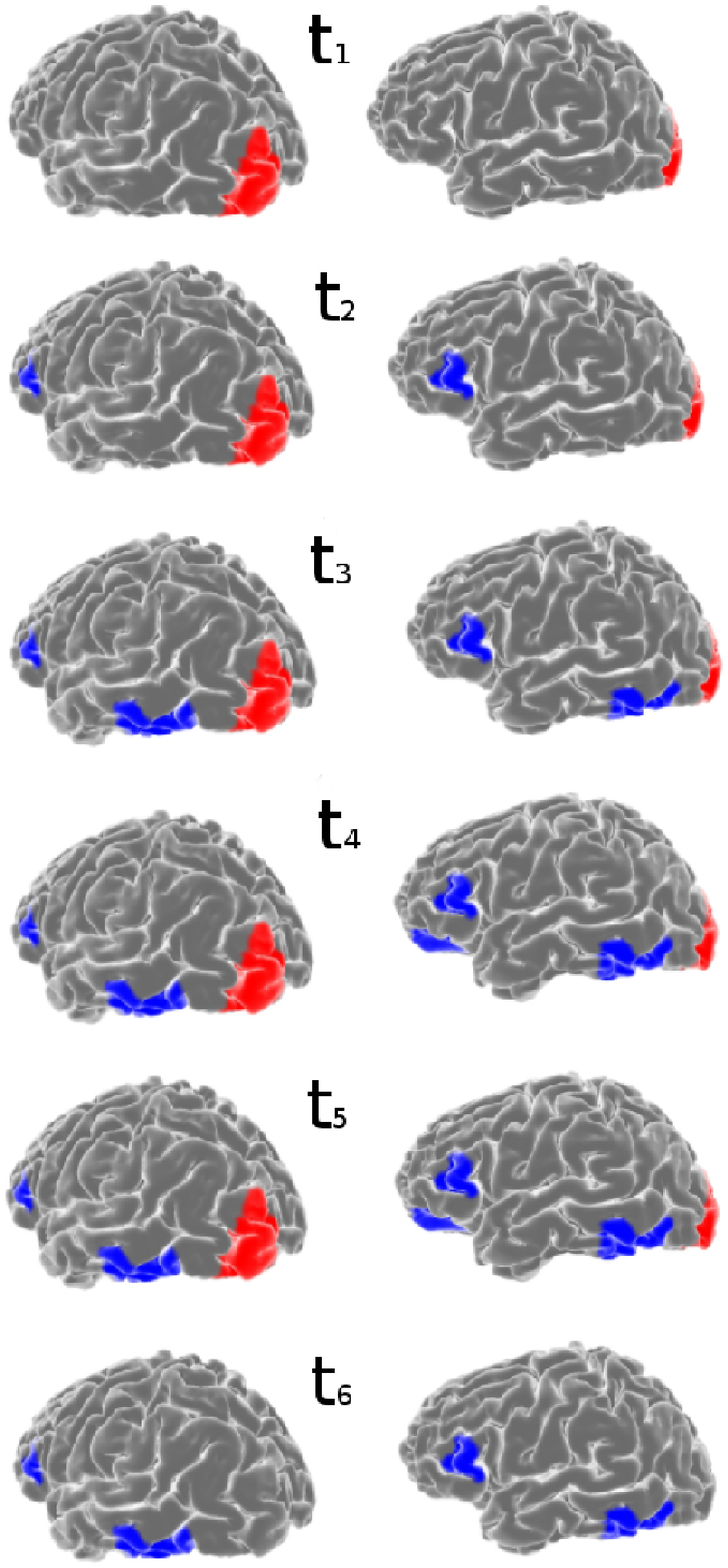}
\end{center}
\caption{Stereotactic Electroencephalographic (SEEG) Data of patient CJ. Left panel: time series of partial seizures recorded with SEEG. The colored bars indicate the EZ (red)and PZ (blue). 
Right panel: On the top, the spatial organization of the EZ and PZ is shown on the left hemisphere, frontal-lateral and lateral view. Below, spatiotemporal activation patterns are plotted at 
different time points of the seizures. The same time points highlighted with black dashed lines in the left panel. lh-SPC, left superior parietal cortex; lh-IPC, left inferior parietal cortex; 
lh-LgG, left lingual gyrus; lh-LOC, left lateral occipital cortex; lh-FuG, left fusiform gyrus; 
lh-PC, left Pericalcarine; lh-ITG, left inferior temporal gyrus. 
}
\label{fig.seizure.data}
\end{figure}

During an epileptic seizure the brain shows an abrupt onset of fast oscillatory activity, which typically propagates to other regions and recruits these as oscillatory subnetworks. 
The seizure originates in the Epileptogenic Zone (EZ) and propagates to the Propagation Zone (PZ) (see Fig. \ref{fig.seizure.data}). The recruitment of the regions in the propagation zone can happen either 
by independent activation of the single areas (Fig. \ref{fig.seizure.data}, time $t_2$), or by activating multiple areas at the same time (Fig. \ref{fig.seizure.data}, times $t_3-t_5$). 
The distinction of EZ and PZ based on the electrographic signatures is non-trivial, in particular, when the spreading continues after the epileptogenic zone has seized its intense activity 
(Fig. \ref{fig.seizure.data}, time $t_6$). In case of asymptomatic seizures, the activity is often limited to the epileptogenic zone and no propagation takes place.

\begin{figure}
\centering
\includegraphics[width=0.27\textwidth]{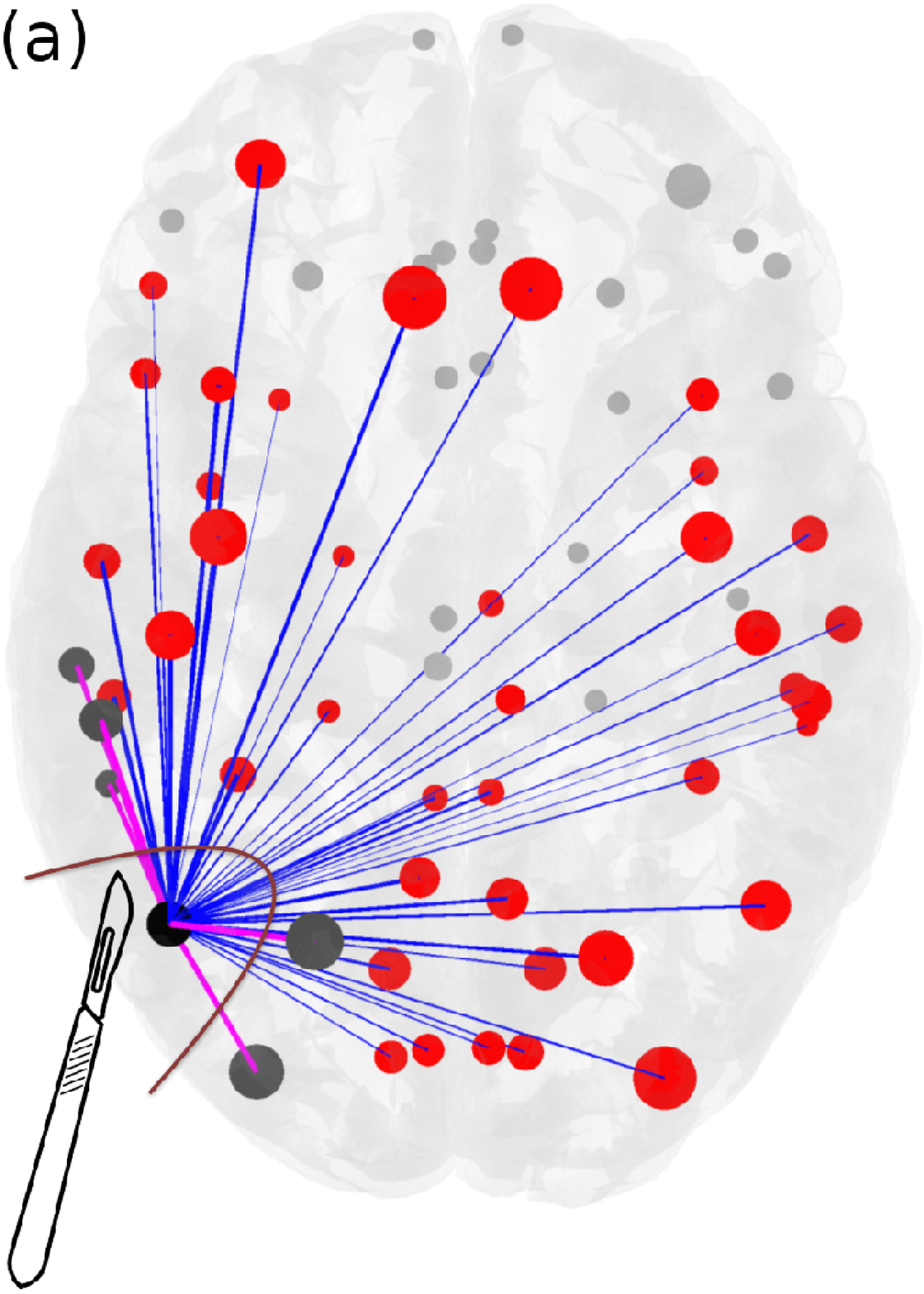}
\includegraphics[width=0.27\textwidth]{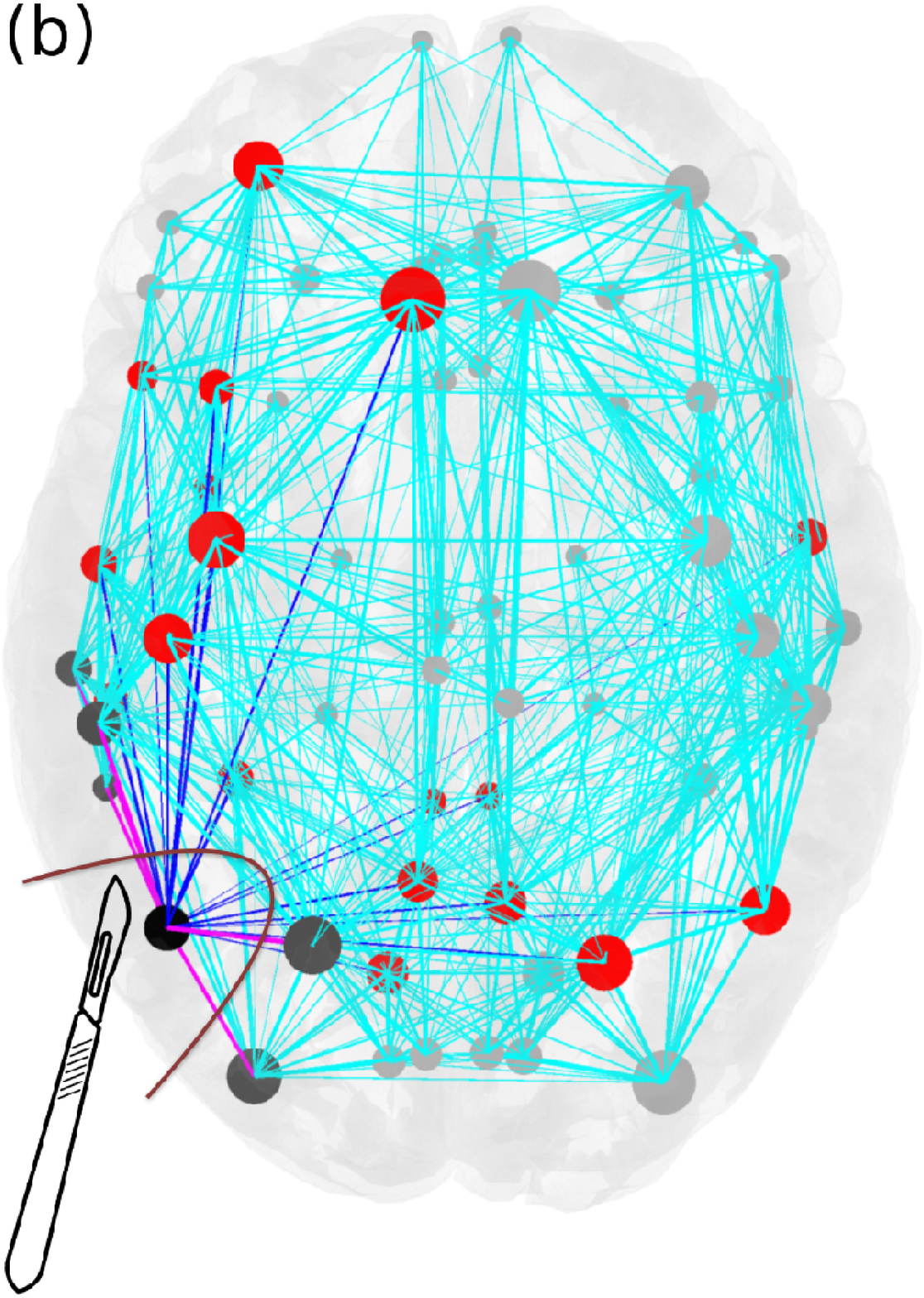}
\caption{Scheme for comparison of standard resection methods, where the entire epileptogenic zone (EZ) is removed during surgical operation, versus lesioning minimal number of links. 
The connectivity matrix is illustrated for an epileptogenic brain with the EZ consisting of one area (black). The outgoing connections of the EZ are blue and pink, 
connecting red and dark grey areas respectively, and they are all removed during the current surgical procedures of disconnecting the EZ. 
Lesioning depicts the minimal number of links that are sufficient to be removed (pink) in order to stop the seizure, versus the total number of outgoing links from
the EZs (blue) that are removed during the resection of an entire EZ. Light blue links added in panel (b) represent the full connectivity of the network. The size of the nodes reflects 
how strongly they are connected, and the width of the links correspond to their weight. Unaffected nodes by any of the resection procedures are light grey and unaffected links are light blue.  }
\label{fig.0}
\end{figure}

\subsection{Comparison among different methodologies} 
\begin{figure}
\begin{center}
\includegraphics[width=0.55\textwidth]{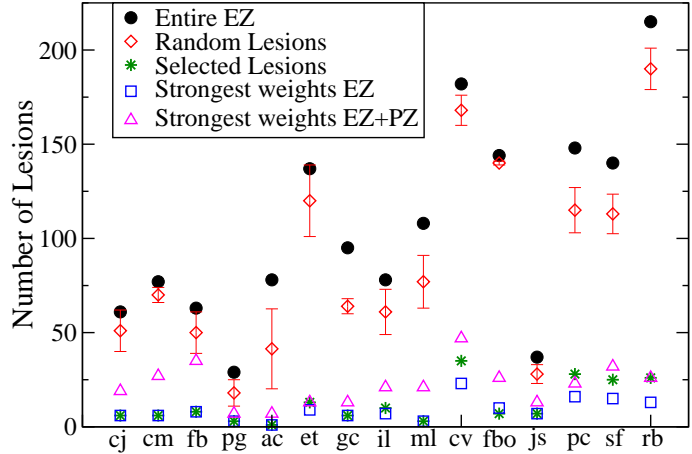}
\end{center}
\caption{Comparison between the results obtained with 5 different lesioning procedures for each of the 15 analyzed patients. In particular it is shown the numbers of links that are cut in order
to stop the seizure propagation, if i) the entire EZ is removed (black dots); ii) random cuts are done (red diamonds); iii) selected lesions are done following
Linear Stability Analysis indications (green stars); iv) the strongest links outgoing the EZ are cut; v) the strongest links outgoing the macroarea (EZ + PZ) are removed. 
Patients are ordered according to the extension of their EZ: the focus is represented as a single node for patients cj-pg; 2 nodes for patients ac-ml; 3 nodes for patients cv and fbo; 4 nodes for 
patients jc-sf; 6 nodes for rb.
}
\label{fig.1}
\end{figure}

Once the EZ is schematized as one (or more) nodes inside the brain network, then
the most common epilepsy surgery actually performed, which is ascribed to the resection of the epileptogenic zone, corresponds to the resection of all the pathways outgoing the 
focal nodes in addition to the removal of the node itself. Fig. \ref{fig.0} shows the connectivity matrix of an epileptogenic brain (light blue links) with the EZ consisting of one area, 
where the set of connections outgoing the focus is highlighted in blue and pink. The targeted disconnection method aims at interrupting only those nerve pathways that allow a seizure 
to propagate and corresponds in this framework to the resections of the few pink links outgoing the EZ. In Fig. \ref{fig.1} we compare the removal of: i) all the links outgoing the epileptogenic zone; 
ii) a set of randomly chosen links outgoing the epileptogenic zone; 
iii) the strongest links outgoing the macro-area composed by epileptogenic and propagation zones; iv) the strongest links outgoing the epileptogenic zone; v) links selected via the Linear 
Stability Analysis procedure. Here strongest links refer to the connections with largest weights. 

The random choice of the links to be cut among the subgraph connecting the EZ with the rest of the network, represent the mathematical
procedure that better approximate the results of standard clinical resections, where all the subgraph is removed, since a big amount of connections needs
to be resected in order to stop the seizure propagation and to prevent the emergence of epileptic symptomatic seizures.
On the other hand, if we restrict our attention to the connections outgoing the region composed of both EZ and PZ, the strength of the links
in this macro-area might be a good indicator to understand the role played by the topology in the spreading of the seizure all along the brain. Actually, 
the mechanism underlying the seizure propagation is more subtle and the topological features of the structural connectivity matrices are not sufficient
themselves to fully explain the recruitment of other areas occurring in the propagation. This means that the number of fundamental links that we identify according to this procedure are 
more than the ones actually responsible for the mechanism. When we apply the same criterion to the subgraph outgoing the EZ only, thus disconnecting the strongest links, 
we are able indeed to design an effective targeted disconnection. Due to the distribution of the weights of the structural connectivity matrices, connections
with biggest weights turn out to be easy to recruit.

The least invasive method identifying the minimal number of pathways, along which the seizure propagates, is the  Linear Stability Analysis, that univocally identifies, 
via the calculation of the stability of the system in presence of perturbations (represented by the seizures), the most unstable
directions along which the recruitment and the seizure spreading take place. These most unstable directions, that are represented by the links connecting
different populations, are often the strongest links outgoing the epileptogenic zone. In particular this is always true when the epileptogenic zone is composed by a single
area, i.e. a single node. However, it is worth noting that purely structural information is not sufficient to predict the propagation and eventual stopping
of the seizure, and it requires the inclusion of the mathematical model to predict correctly the least invasive intervention. To demonstrate this, we compute various network measures, 
including efficiency, strength, betweeness, centrality and clustering (see Figs. S2-S4), which show that neither the epileptogenic zone 
nor the propagation zone are characterized by outstanding values of these indicators. In general, epileptogenic and propagation zones have intermediate values of all the graph metrics. 
On the other hand, the Linear Stability Analysis offers a way to calculate the stability of the system against perturbations and to identify the directions along which the system is the least stable. 
Thus, the lesion procedure, identified with the Linear Stability Analysis, is to be preferred to the analysis of the strongest links only, since it is able to give insights on the dynamics of 
the system and its criticalities, even though its predictions are less effective for extended EZ.

\subsection{Detailed analysis of a single patient}

\begin{figure}
\begin{center}
\includegraphics[width=4.5cm,height=6.65cm]{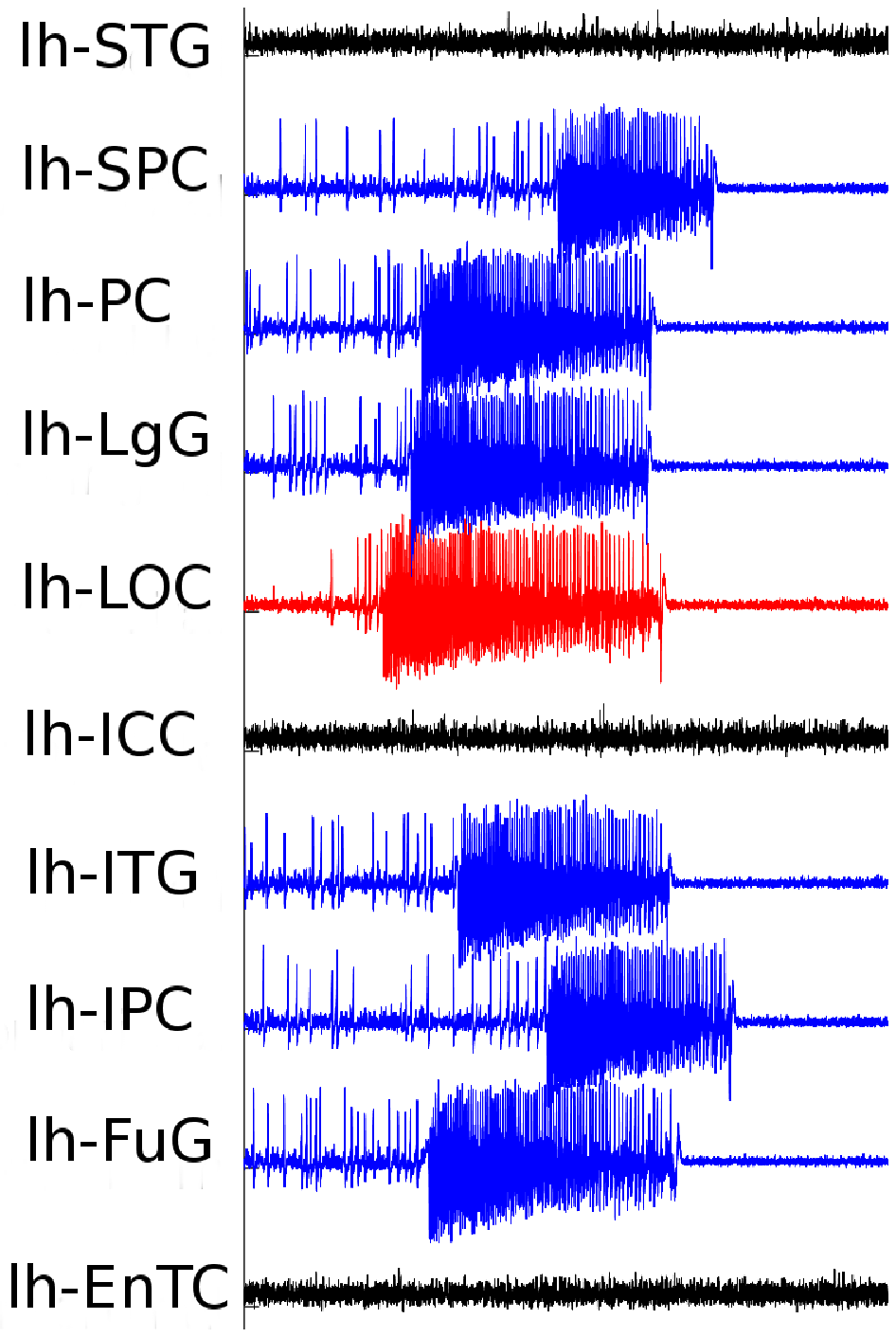}
\includegraphics[width=3.2cm,height=6.65cm]{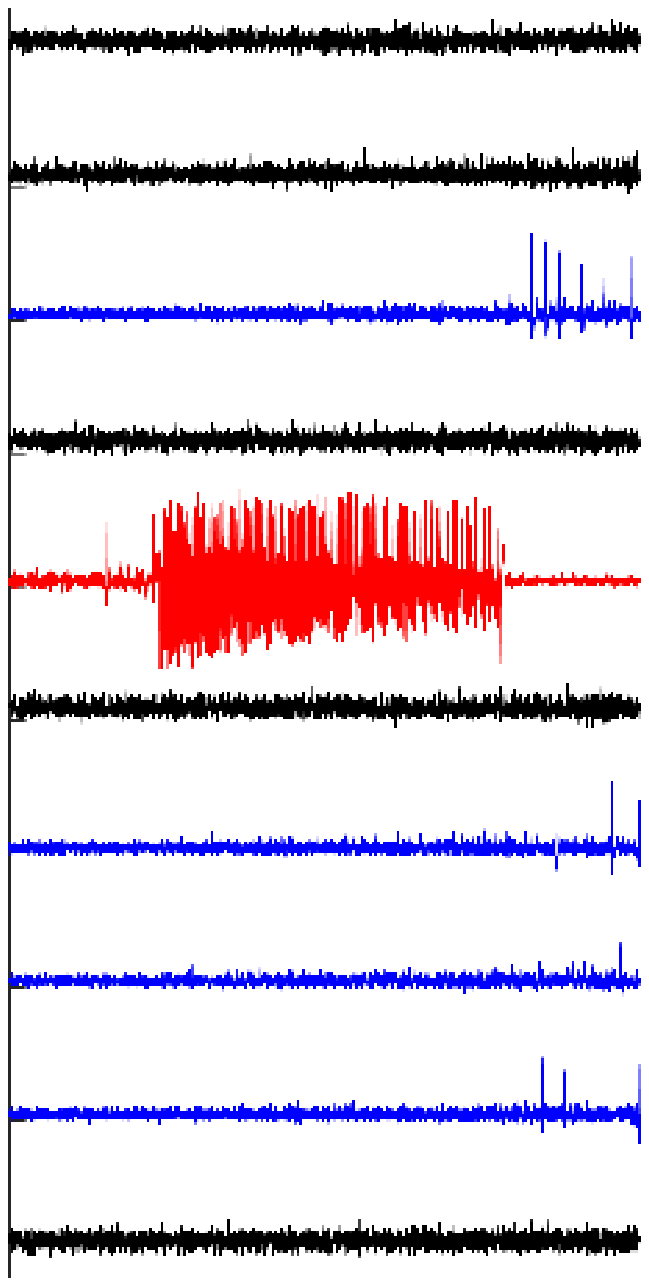}
\includegraphics[width=0.5\textwidth]{RasterPlotCJ.eps}
\end{center}
\caption{ Analysis of patient \textbf{CJ}. Clinical history: occipital epilepsy type (left size). EZ region: Lateral occipital cortex. 
PZ prediction: Fusiform gyrus, Superior parietal cortex, Inferior temporal gyrus, Inferior parietal cortex, Pericalcarine, Lingual gyrus.
PZ clinical prediction: Inferior parietal cortex, Superior parietal cortex.
Lesions: links between regions LOC-FuG, LOC-SPC, LOC-ITG, LOC-IPC, LOC-PC, LOC-LgG  must be cut in order to stop the seizure
(lesions are performed in correspondence of the time identified by the dashed blue line).
Upper panel: Time series generated by the implemented brain network model with the connectome of patient CJ. On the left the PZ (blue curves) is recruited
immediately after the seizure (red curve) is emitted; on the right the recruitment is no more possible after the targeted disconnection is performed.
Lower panel: Seizure events as a function of time. Green dashed line corresponds to the time at which lesions occur.
The blue ellipse highlights the moment in time at which the PZ is recruited after the seizure emitted by the EZ.
}
\label{fig.2}
\end{figure}

From the computational point of view, the dynamics of each node of the network parcellation is given by a neural mass model able to reproduce the temporal seizure dynamics and to switch 
autonomously between interictal and ictal states \cite{Jirsa2014}. Nodes are connected together via a permittivity coupling
acting on a slow time-scale, that is on the time-scale of seconds, which is sufficient to describe the recruitment of other brain regions \cite{Proix2014}. The self-emergent dynamics
of this system, once the connectivity matrix of patient CJ is chosen, is shown in Fig. \ref{fig.2} (upper panels). The EZ is the Lateral occipital cortex (LOC), placed on the left hemisphere.
The seizure onset is followed by the recruitment and successive seizure emissions of the PZ: Fusiform gyrus (FuG), Superior parietal cortex (SPC), Inferior temporal gyrus (ITG), Inferior parietal cortex (IPC), 
Pericalcarine (PC), Lingual gyrus (LgG), all located in the left hemisphere (upper left panel). Once the pathways to be resected are chosen via the Linear Stability Analysis suggestions and the 
targeted disconnection is performed, the recruitment process is not anymore possible in the system even though the EZ is still triggering seizures (upper right panel).

A different visualization of the seizures onsets, taking place in the network at different times, is shown in Fig.  \ref{fig.2} (lower panel): each seizure emitted by a single node at a specific time is 
identified with a black (red) dot, as in a spike raster plot, which is a typical representation of neural spike-timing activity. In this way it is possible to have an overview of the entire network for 
long time periods. In particular the epileptogenic zone is represented in the raster plot as node 21 and its emitted seizures are identified with red dots for consistency with the upper panel. 
EZ is able to trigger seizure autonomously and, after a small time interval, to recruit a small number of regions: the recruitment event highlighted in the blue circle is the same as shown in the upper left panel. 
After a small delay, suddenly, the seizure is no more confined to a small number of regions, but it is able to propagate along the 
entire network, and all regions start emitting seizures synchronously (as in a bursting activity). This mechanism is self-sustained unless selected lesions are performed.
In particular, at the time corresponding to the blue dashed line in Fig.  \ref{fig.2} (lower panel), six lesions are performed on the structural connectivity and the seizure is 
no longer able to recruit any sub-network. The ongoing dynamics after the disconnection is performed, is the same shown in Fig.  \ref{fig.2} (upper right panel).
The lesions correspond to cutting the connections between the EZ and the nodes in the PZ that are involved in the initial recruitment phase and are identified by calculating 
the maximal eigenvector of the Jacobian matrix (see SI).

\begin{figure}
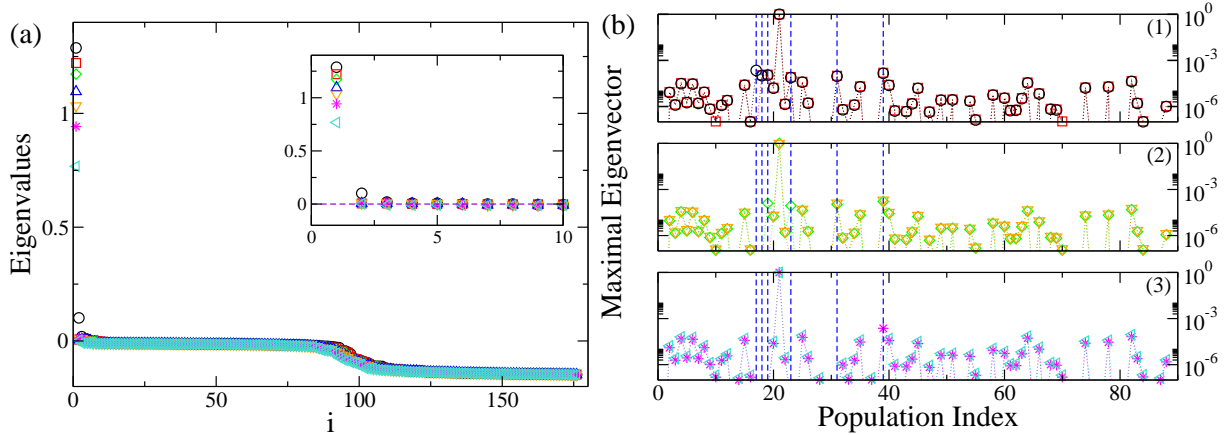

\begin{center}
\includegraphics[width=0.47\textwidth]{Spettro.eps}
\includegraphics[width=0.5\textwidth]{AutovettoreMax.eps}
\end{center}
\caption{Analysis of patient \textbf{CJ}. Panel (a): Eigenvalues of the system  when the original structural connectivity is considered as a network structure (black circles). 
The index $i$ on the x-axis represents the eigenvalues index (for more details see SI).
Red squares represent the eigenvalues of the networks when the link between the populations (nodes) LOC-FuG (21-17) is removed. 
Green diamonds represent the eigenvalues of the networks when the links between the populations (nodes) LOC-FuG, LOC-SPC (21-17, 21-39) are removed. Blue triangles
represent the eigenvalues of the networks when 3 connections between the regions (nodes) LOC-FuG, LOC-SPC, LOC-ITG (21-17, 21-39, 21-19) are removed.
Orange triangles represent the activity when 4 connections between the regions (nodes) LOC-FuG, LOC-SPC, LOC-ITG, LOC-IPC (21-17, 21-39, 21-19, 21-18) are removed. 
Magenta stars: links between the regions (nodes) LOC-FuG, LOC-SPC, LOC-ITG, LOC-IPC, LOC-PC (21-17, 21-39, 21-19, 21-18, 21-31) are removed.
Turquoise triangles: links between the populations (nodes) LOC-FuG, LOC-SPC, LOC-ITG, LOC-IPC, LOC-PC, LOC-LgG (21-17, 21-39, 21-19, 21-18, 21-31, 21-23) are removed.
In the inset is shown an enlargement of the first, positive part of the spectra.
Panel (b): Maximal eigenvectors calculated when the original structural matrix is considered as network structure and
when successive links are removed. Symbols and the color code are the same as in panel (a).
The blue lines represent the most unstable localized nodes to which are associated the most unstable directions.
}
\label{fig.3}
\end{figure}

The dynamics presented in Fig. \ref{fig.2}, can be also characterized in terms of the eigenvalues shown in Fig.  \ref{fig.3} (a).
In particular, the dynamics describing the seizures of patient CJ, is characterized by 2 positive eigenvalues, thus meaning that the 
system in unstable. The eigenspectrum changes if some topological modifications (e.g. lesions) are implemented into the original structural matrix.
In particular the number of positive eigenvalues diminishes by each further removal of the links, following the Linear Stability Analysis procedure. 
It is important to notice that the necessary links to be cut in order to stop the seizure are those between the EZ (node 21) and the first six
most unstable nodes. Once these links are removed the system is still unstable, because the epileptogenic zone is still active,
but the epileptogenic activity of this area is no more able to recruit other zones.

In Fig. \ref{fig.3} (b), the procedure for identifying the most unstable nodes is described in more details. Firstly, we show the maximal eigenvector calculated 
for different network configurations. The maximal eigenvector is associated with the maximum positive eigenvalue resulting from the calculation of the Jacobian matrix.
In this calculation is taken into account also the connectivity of the specific patient.
Therefore, when the original structural connectivity matrix is considered as network structure, the largest elements of the eigenvector correspond to some populations
in the brain parcellation and the most unstable directions turn out to be the connections between these populations and the EZ. The largest elements of the eigenvector
are indicated by blue dashed lines in Fig.  \ref{fig.3} (b); the biggest one (non indicated with a dashed line) is the EZ. In particular, if the link between the epileptogenic population 
and the second largest element (corresponding to FuG - node 17) is cut (sub-panel (1), red squares), this direction is no longer unstable and the node looses its importance,  
as element of the maximal eigenvector, not having anymore a relevant amplitude. In particular the entry corresponding to the node, whose connection has been resected, is now characterized by a
very small absolute value, while other directions may be enhanced. 
The other largest elements of the vector (indicated by blue lines) still constitute the set of most unstable directions, while this is no longer true for the resected connection LOC-FuG (21-17), 
related to the entry number 17 of the maximum eigenvector. This discussion applies to the first 6 biggest elements of the eigenvector.
Once all the 6 links are cut (see sub-panel (3), turquoise triangles), the eigenvector is no more localized in specific zones. The EZ  
is the only one left significantly larger than zero, but it is no longer contributing to the instability of the system in a relevant way.

\subsection{Systematic analysis} 
\begin{figure}
\begin{center}
 \includegraphics[width=0.45\textwidth]{SystematicAnalysis.eps}
\includegraphics[width=0.5\textwidth]{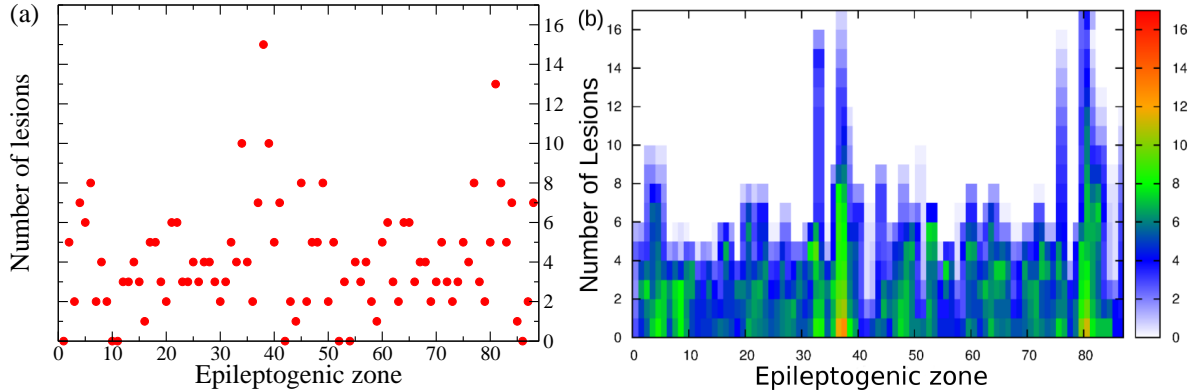}
\end{center}
\caption{Panel (a): Number of lesions necessary to stop seizure propagation as a function of the epileptogenic region.
Panel (b): Dependence of the PZ size on each lesion, as a function of the EZ. The color code indicates
the number of regions belonging to the PZ, from 0 (white pixel) to 17 (red pixel).
}
\label{fig.4}
\end{figure}

The former analysis has been done on the basis of the medical doctors prediction for the EZ. However it is worth doing a systematic analysis of the network dynamics, 
by supposing that the EZ is placed in every possible node of the connectome. In particular, by taking into account the structural connectivity matrix of patient CJ and by systematically 
placing the EZ in different nodes, we investigate the dynamics of the full network for all possible epileptogenic nodes. For every shuffled position of the EZ, the Linear Stability Analysis is
calculated, henceforth the eigenvalues and eigenvectors of the system are obtained for all these configurations. Even tough the results depend on the position of the EZ, it is possible to demonstrate that, 
in general, by performing a virtual resection of a small number of pathways, up to a maximum of 15, the seizure propagation can be stopped for any initial EZ considered (see Fig. \ref{fig.4}(a)).
This proves that the performed lesioning procedure is independent on the chosen particular case and it can be validated for general EZ locations.

The lesioning procedure for a given structural connectivity network directly affects the size of the PZ. As expected, it shrinks when increasing the number of removed links.
This is a natural consequence of the fact that the number of possible unstable directions decreases for every resection, and the number of regions that are likely to be recruited vanishes. 
More specifically, the PZ generally shrinks to 0-2 areas, after 5-6 performed disconnections (see Fig. \ref{fig.4}(b)). 
In most of the cases this is sufficient to alter the seizure spreading, since a small PZ, made of 2 populations maximum, is not able to convey the spreading: the
most unstable directions are no more accessible by the system. In this case the system undergoes a so-called asymptomatic seizure. 

\begin{figure}
\begin{center}
\includegraphics[width=0.6\textwidth]{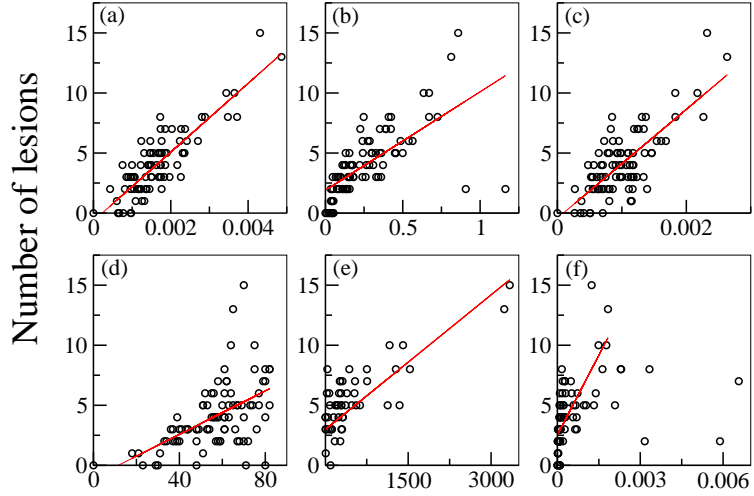}
\end{center}
\caption{Number of lesions needed to stop the propagation for different graph measures of each EZ, calculated for the connectivity matrix of the patient CJ. Graph metrics reported on the x-axis are: 
(a) efficiency; (b) strength; (c) clustering; (d) degree; (e) betweenness; (f) centrality. 
The red line in each panel shows the correlation among the data, and the estimated Pearson correlation coefficient is : (a) 0.8615377; (b) 0.6735522; (c) 0.765353;
(d) 0.540695; (e) 0.7852488; (f) 0.7165724. All p values of statistical significance are below $10^{-6}$.
}
\label{fig.cj.topology}
\end{figure}

A topological analysis \textit{a posteriori} on the dependence of the partial targeted disconnection on the different graph connectivity measures, highlights
the significance of these quantities. Namely, starting from the results reported in Fig. \ref{fig.4}(a), we have further calculated the dependence of the disconnection procedure on efficiency, 
strength, clustering, degree, betweenness, centrality (see Fig. \ref{fig.cj.topology}). In particular these topological metrics have different values for different choices of the EZ.
The results indicate that the number of needed topological interventions on the network are strongly proportional to the efficiency, clustering and betweenness, whereas, while still present, 
this relationship is less pronounced for the strength, degree and centrality. This means that controlling the seizure spreading is more difficult when
the involved EZ belongs to a dense, high-clustered neighborhood. In this situation each node involved can easily transmit information among the others:  
a clique of regions that rapidly communicate by using a dense subgraph plays a big role in controlling the network and enhancing the seizure propagation. Once the clique controls 
the dynamics, it is necessary to cut more links responsible for the information flow to propagate. On the contrary, the shortage of effective communication networks among the pathological areas 
facilitates the control of the network, and in this case the goal can be achieved with a much smaller number of topological interventions.

The importance of separately considering the structural connectivity matrices of single patients, for better designing the partial targeted disconnection,
can be illustrated comparing the previous results with the ones obtained by performing the Linear Stability Analysis procedure on an average connectome.
The latter is constructed by averaging the 15 structural connectivity matrices of the considered patients (via an algebraic mean).
It turns out that the analysis on the average connectome is much less predictive and that the individual variability has clear effects upon the outcome of the prediction results, 
and therefore, of possible therapies and treatment approaches. In particular, the resulting average connectome has different topological characteristics with respect to the matrices of the single patients:
the degree of each node is larger and in general, it is more dense and homogeneous. If we select as EZ, the area suggested by the clinical prediction (i. e. node 21, lh-LOC
for the patient CJ) and we calculate the maximum eigenvector, it turns out to be no longer localized, like the one shown in Fig. \ref{fig.4} (b). 
Now the eigenvector has a distributed structure, with many elements having a considerably large absolute value (i. e. above $10^{-5}$). 
Hence, it is more difficult to individuate a small number of nodes that actively participate to the dynamics and that are overwhelmingly involved in the seizure propagation. 
On the contrary, many more nodes are now involved in the propagation, and many more directions are viable for the spreading of the pathological activity.
A comparative analysis of the number of links that must be cut for stopping the seizure propagation, obtained by following different procedures, is reported in  Table 1.

\begin{table}[h]
\begin{ruledtabular}
   \begin{tabular}{ l l }
   \hline
    \textbf{Type of analysis} & \textbf{Lesions}\\ \hline
    Linear Stability Analysis (LSA) on the \\ patient's structural connectivity matrix (PSCM) &  6   \\ \hline
    Strongest Links in the PZ (on the PSCM) & 22  \\ \hline
    Random Cuts (on the PSCM) &  51   \\ \hline
    All the outgoing links of the EZ (on the PSCM) & 61 \\ \hline
    LSA on the Average Connectome & 66   \\ \hline
    Random Cuts on the Average Connectome & 69   \\ \hline
    All the outgoing links of the EZ \\ on the Average Connectome & 71 \\ \hline
    \end{tabular}
    \end{ruledtabular}
    \label{tab:1}
    \caption{Comparison of the minimum necessary number of lesions for stopping the seizure propagation for different lesioning procedures.}
\end{table}

While it is sufficient to cut only 6 links detected via the Linear Stability Analysis procedure on single patient's connectome, it is necessary to remove up to 66 links 
before the seizure stops propagating on the averaged connectome, even thought the same procedure have been used. The latter number is comparable with the results obtained by random removal of the links 
belonging to the subgraph of the EZ in the patient's connectome, and hence it is no better than a ``blind'' resection.

\section{Discussion}

Connectivity is fundamental to the transmission of information through a network, let it be pathological or physiological. With regard to the mechanisms underlying propagation in networks, 
several mechanisms have been hypothesized. A basic form of propagation is energy dissipation \cite{Spiegler2016}, which depends on the physical properties of the network such as the local 
response of the tissue and its connectivity. Following a local stimulation, the brain network shows initially the well-known local propagation via the intracortical connections within the 
grey matter, but then the spatiotemporal organization after 300 msec is dominated by the large-scale white matter connection topology \cite{Spiegler2016}. As this process is independent of 
any representation of function, propagation within cognitive networks rely at least implicitly on some notion of coding, that renders an activation profile functionally meaningful. 
For instance, increased interareal gamma-band coherence and phase synchronization have been hypothesized as markers of selective attention allowing different input information streams to be 
routed through brain networks according to task relevance \cite{Palmigiano2017, Gregoriou2009, Grothe2012}. In the diseased brain the propagation of this information is disturbed. 
Seizure propagation, in particular, is generally based on the concept of abnormalities of synchronizability: i) seizure evolution is driven by the strong synchronizing activity of the EZ which  
is able to guide the surrounding tissue \cite{abnormalEZ}; ii) the surrounding tissue has a diminished ability to contain and regulate abnormal activity, and it allows the seizure to 
propagate \cite{abnormalPZ}; iii) synchronizing and desynchronizing nodes operate antagonistically, such that synchronizability increases and dynamical processes diffuse easily through the 
network when synchronizing nodes exert greater push than desynchronizing nodes \cite{bassett2016}. 

In all these cases of propagation, the frame, in which the activity evolves, is defined by the energy dissipation properties of the network and can be controlled and rerouted through network rewiring, 
of which the disconnection type is the simplest to realize, even though in clinical practice it has not found many applications so far. The few existing applications exclusively rely on a complete 
disconnection such as corpus callosotomy or hemispherotomy, since in absence of understanding the network effects upon propagation, 
more subtle and minimally invasive interventions are impossible. As a paradigmatic type of pathological propagation, we considered seizure spread. Our in-silico approach allowed an exhaustive search, 
which would have been impossible in vivo. The model predicts the activity propagation of a seizure originating in a certain epileptogenic zone and to identify the most unstable pathways that 
support and allow the propagation. In other words, activity propagation cannot be disjointed by the network topology outside the epileptogenic zone and by the topological features linked to the propagation 
zone that both contribute to the propagation. We have demonstrated that the energy dissipation of seizure propagation may be controlled by performing a Linear Stability Analysis of the system, 
thus allowing identification of the properties and the response of the system undergoing a perturbation (such as a seizure emission).
In this framework, the most unstable directions that carry on the spreading may be identified and, by studying the effect of lesions on the dynamics of the simulated brain processes, 
it is possible to visualize when the seizure stops propagating, and the recruitment of the PZ does not occur anymore. Other groups have already studied the effects of removing brain regions 
and connections on the dynamics of simulated network processes \cite{virtual_resection}, but in all these works the interest is mainly focused on understanding the effect of damages on the brain activity, 
instead of using selected lesions as the least invasive treatment as possible in an eventual surgical operation. Moreover, our approach is entirely based on structural data, 
which allows the creation of a brain model based on purely non-invasive data prior to any surgery or exploration via intracranial measurements, whereas other techniques rely on invasive functional 
data focused on the estimation of the epileptogenic zone \cite{Burns2014} and resective surgery \cite{Taylor2014, Sinha2017}. Disconnection approaches thus are significantly 
less invasive and bear the potential for general network control and applications beyond seizure suppression. 

Our approach generalizes to oscillatory activity propagation so far, that it does not necessitate the presence of the hallmark of epilepsy, that is an onset and offset of high-frequency oscillatory activity. 
It thus can be applied in the future to understand the directionality of state changes and manipulate the activity in the network. On the other hand, concerning the applicability to 
the seizure propagation problem, it may be determinant to design a new surgical approach, providing support and improving the resection surgical procedure. 
A less invasive and more effective surgical approach would make the difference in short and long-term outcomes and in improving the quality of life of the patients. 
Moreover, and eventual persistence of seizures after surgery would not inhibit a second clinical trial for patients, since the procedure does not consist in removing consistent parts of the brain, 
that on the contrary, would impair the physical abilities of the patient.


\section{Methods}

\subsection{Patient selection and data acquisition}

We selected 15 drug-resistant patients (9 females, 6 males, mean age 33.4, range 22-56) with different types of
partial epilepsy accounting for different EZ localizations. All patients underwent a presurgical 
evaluation (Supplementary Table 1). The first phase in the evaluation of each patient is not
invasive and comprises of the patient clinical record, neurological examinations, positron emission
tomography (PET), and electroencephalography (EEG) along with video monitoring. T1 weighted
anatomical images (MPRAGE sequence, TR=1900 ms, TE=2.19 ms, 1.0 x 1.0 x 1.0 mm, 208 slices) and
diffusion MRI images (DTI-MR sequence, angular gradient set of 64 directions, TR=10.7 s, TE=95 ms, 2.0
x 2.0 x2.0 mm, 70 slices, b weighting of 1000 $s/mm^2$) were acquired on a Siemens Magnetom Verio
3T MR-scanner. From the gathered data clinicians conclude potential epileptogenic zones (EZ). Further
elaboration on the EZ are done in the second phase, which is invasive and consists of placement of
stereotactic EEG (SEEG) electrodes in or close to the suspected regions. These electrodes have 10 to 15
contacts that are 1.5 mm apart. Each contact is 2 mm of length and 0.8 mm in diameter. The SEEG was
recorded by a 128 channel Deltamed$^{TM}$ system using a 256 Hz sampling rate. The SEEG recordings were
band-pass filtered between 0.16 and 97 Hz by a hardware filter. All the chosen patients showed seizures
in the SEEG starting in one or several localized areas, that is, the EZ before recruiting distant regions,
i.e. the PZ. The position of the electrodes was pinned down by performing a
computerized tomography (CT) scan or a MRI after implanting the electrodes.

\subsection{Clinical definition of the Propagation Zone}

The propagation was defined by two different methods. The first method is the subjective evaluation of
clinicians based on the different measurement modalities (EEG and SEEG) gathered throughout the two-step 
procedure (non-invasive and invasive). The second method is objective in the sense that it is solely
based on the SEEG measurements. For each patient, all seizures were isolated in the SEEG time series. The bipolar SEEG
was considered (between pairs of electrode contacts) and filtered between 1-50 Hz using a Butterworth
band-pass filter. A contact was considered to be in the PZ if its signal energy was responsible for at least
30$\%$ of the maximum signal energy over the contacts, and was not in the EZ. The corresponding region
was then assigned to the PZ.

\subsection{Network measures}

Topological properties of a network can be examined by using different graph measures that are provided by the general framework of the graph theory.
These graph metrics can be classified in measures that are covering three main aspects of the topology: segregation, integration and centrality.
The segregation  accounts for the specialized processes that occur inside a restricted group of brain regions, usually densely connected.
Among the possible measures of segregation, we have considered the \textit{clustering coefficient}, which gives the fraction of triangles around a node and it is 
equivalent to the fraction of node's neighbors that are neighbors of each other as well. This is an important measure of segregation, 
since the presence of a dense neighborhood around a node is fundamental to the generation of clusters and cliques that are likely to share specialized information.

The measures of integration refer to the capacity of the network to rapidly combine specialized information from not nearby, distributed regions.
Integration measures are based on the concept of communication paths and path lengths, which estimate the unique sequence of nodes and links that are able to carry the transmission 
flow of information between pairs of brain regions. In particular the path length measures the sum of the edge weights in a weighted graph (as the one that we have considered)
and the shortest path length suggest a stronger potential for integration between different brain regions. At the global level it is moreover possible to define a ``characteristic
path length'' of the network, which is calculated as the average inverse shortest path length between the all pairs of nodes in the network.
Among the possible measures of integration, as the most meaningful for our structural connectivity matrices we have chosen the \textit{efficiency}. It is defined as the the average inverse 
shortest path length in the network, and is inversely related to the characteristic path length.

Centrality refers to the importance of network nodes and edges for the network functioning. The most intuitive  index of centrality is the \textit{node degree}, which
gives the number of links connected to the node; for this measure connection weights are ignored in calculations. The \textit{node strength} is the sum of weights of links connected to the node,
while node \textit{betweenness centrality} is the fraction of all shortest paths in the network that contain a given node. Nodes with high values of betweenness centrality 
participate in a large number of shortest paths. Finally, \textit{closeness centrality} is the inverse of the average shortest path length from one node to all other nodes
in the network.

\begin{acknowledgments}
 S.0. acknowledge Benjamin Lindner, Dionysios Perdikis, Timoth\'{e}e Proix, Marmaduke Woodman for useful discussions.
This work was partially supported by The Short Term Mobility Programm founded by the Italian National Research Counsil
(protocol number 0057676, date 31/08/2015) and by the Project Epinext - 
F\'{e}d\'{e}ration Hospitalo-Universitaire (FHU) ``Epilepsy and Disorders of Neuronal Excitability''.
\end{acknowledgments}

\section*{Author Contributions}
Conceived and designed the experiments: SO VJ. Performed the experiments and analyzed the data: SO SP. Provided the data: MG FB. Wrote the paper: SO SP MG FB VJ

\section*{Additional Information}
Competing financial interests: The authors declare no competing financial interests. \\
Non-financial competing interests: The authors declare no competing non-financial interests.


\begin{thebibliography}{1000}



\bibitem{Kirst2016}
Kirst, C., Timme, M., Battaglia, D., \textit{Dynamic information routing in complex networks.}, Nature Communications 7, 11061 (2016).

\bibitem{Palmigiano2017}
Palmigiano, A., Geisel, T., Wolf, F., Battaglia, D., \textit{Flexible information routing by transient synchrony}, Nature Neuroscience (2017).


\bibitem{Roelfsema} 
Lorteije, J. A., Zylberberg, A., Ouellette, B. G., De Zeeuw, C. I., Sigman, M., and Roelfsema, P. R.,  \textit{The formation of hierarchical decisions in the visual cortex}. Neuron, 87(6), 1344-1356 (2015).

\bibitem{Alexa_Riehle}
Ponce-Alvarez, A., N\'{a}cher, V., Luna, R., Riehle, A. and Romo, R., \textit{Dynamics of cortical neuronal ensembles transit from decision making to storage for later report}. 
J. Neurosci., 32(35), 11956-11969 (2012).

\bibitem{Daffertshofer}
van Wijk, B.C., Beek, P.J. and Daffertshofer, A., \textit{Neural synchrony within the motor system: what have we learned so far?}, Front. Hum. Neurosci., 6:252, 
(2012).
de Vries, I.E., Daffertshofer, A., Stegeman, D.F. and Boonstra, T.W., \textit{Functional connectivity in the neuromuscular system underlying bimanual coordination}. J. Neurophys., 116(6), 2576-2585, (2016).
	
\bibitem{Barry_Horwitz}
Horwitz, B. and Braun, A.R., \textit{Brain network interactions in auditory, visual and linguistic processing}. Brain Lang., 89(2), 377-384 (2004).

\bibitem{Sleimen-Malkoun2015}
Sleimen-Malkoun, R., Perdikis, D., M\"{u}ller, V., Blanc, J.L., Huys, R., Temprado, J.J. and Jirsa, V.K., \textit{Brain dynamics of aging: multiscale variability of EEG signals at rest and during an 
auditory oddball task}. Eneuro, 2(3), pp.ENEURO-0067 (2015).

\bibitem{Battaglia2018}
Battaglia, D., Boudou, T., Hansen, E. C. A., Chettouf, S., Daffertshofer, A., McIntosh, A. R., Zimmermann, J., Ritter, P., and Jirsa, V.,  \textit{Functional Connectivity Dynamics of the Resting State across 
the Human Adult Lifespan}. bioarXiv doi: https://doi.org/10.1101/107243. 

\bibitem{Destexhe2016}
Dehghani, N., Peyrache, A., Telenczuk, B., Le Van Quyen, M., Halgren, E., Cash, S.S., Hatsopoulos, N.G. and Destexhe, A., \textit{Dynamic balance of excitation and inhibition in human and monkey neocortex},
Scientific reports, 6 (2016).

\bibitem{Isomura2008}
Isomura Y., Fujiwara-Tsukamoto Y., Takada M., \textit{A network mechanism underlying hippocampal seizurelike synchronous oscillations}, Neurosci Res. 61 (3), 227-233. (2008).

\bibitem{Wendling2002}
Wendling F., Bartolomei F., Bellanger J. J., Chauvel P., \textit{Epileptic fast activity can be explained by a model of impaired GABAergic dendritic inhibition},
Eur J Neurosci. 15(9), 1499-1508 (2002).

\bibitem{Bernard2000}
Bernard C., Cossart R., Hirsch J. C., Esclapez M., Ben-Ari Y., \textit{What is GABAergic inhibition? How is it modified in epilepsy?}, Epilepsia 41(s6) (2000).

\bibitem{Regis2017}
R\'{e}gis, J., Helen Cross, J. and Kerrigan, J. F., \textit{Achieving a cure for hypothalamic hamartomas: a Sisyphean quest?}, Epilepsia 58 7-11 (2017). 

\bibitem{Ngugi2010}
Ngugi, A.K., Bottomley, C., Kleinschmidt, I., Sander, J.W. and Newton, C.R., \textit{Estimation of the burden of active and life?time epilepsy: a meta-analytic approach}, Epilepsia, 51(5), pp.883-890 (2010).

\bibitem{Blake2000}
Blake, R.V., Wroe, S.J., Breen, E.K. and McCarthy, R.A., \textit{Accelerated forgetting in patients with epilepsy: evidence for an impairment in memory consolidation}, Brain, 123(3), pp.472-483 (2000).

\bibitem{Wang2011}
Wang, Z., Lu, G., Zhang, Z., Zhong, Y., Jiao, Q., Zhang, Z., Tan, Q., Tian, L., Chen, G., Liao, W. and Li, K., \textit{Altered resting state networks in epileptic patients with generalized tonic?clonic seizures},
Brain research, 1374, pp.134-141 (2011).

\bibitem{Schwab1939}
Schwab, R.S. and Cobb, S., \textit{Simultaneous electromyograms and electroencephalograms in paralysis agitans}, Journal of neurophysiology, 2(1), pp.36-41 (1939).

\bibitem{Talairach1966}
Talairach, J. and Bancaud, J., \textit{Lesion," irritative" zone and epileptogenic focus}, Stereotactic and Functional Neurosurgery 27(1-3), 91-94 (1966).

\bibitem{Spencer2002}
Spencer, S.S., \textit{Neural networks in human epilepsy: evidence of and implications for treatment}, Epilepsia 43(3), 219-227 (2002).

\bibitem{Guye2008}
Guye, M., Bartolomei, F. and Ranjeva, J.P., \textit{Imaging structural and functional connectivity: towards a unified definition of human brain organization?}, Current opinion in neurology 21(4), 393-403 (2008).

\bibitem{Wendling2010}
Wendling, F., Chauvel, P., Biraben, A. and Bartolomei, F., \textit{From intracerebral EEG signals to brain connectivity: identification of epileptogenic networks in partial epilepsy},
Frontiers in systems neuroscience, 4 (2010).

\bibitem{Laufs2012}
Laufs, H., \textit{Functional imaging of seizures and epilepsy: evolution from zones to networks}, Current opinion in neurology 25(2), 194-200 (2012).

\bibitem{Bartolomei2013}
Bartolomei, F., Bettus, G., Stam, C.J. and Guye, M., \textit{Interictal network properties in mesial temporal lobe epilepsy: a graph theoretical study from intracerebral recordings},
Clinical Neurophysiology 124(12), 2345-2353 (2013).

\bibitem{daSilva2013}
Stefan, H. and da Silva, F.H.L., \textit{Epileptic neuronal networks: methods of identification and clinical relevance}, Frontiers in neurology 4 (2013).

\bibitem{Kwan2011}
Kwan, P., Schachter, S.C. and Brodie, M.J., \textit{Drug-resistant epilepsy}, New England Journal of Medicine 365(10), 919-926 (2011).

\bibitem{Fisher2012}
Fisher, R.S., \textit{Therapeutic devices for epilepsy}, Annals of neurology 71(2), 157-168 (2012).

\bibitem{Fisher2014}
Fisher, R.S. and Velasco, A.L., \textit{Electrical brain stimulation for epilepsy}, Nature Reviews Neurology 10(5), 261-270 (2014).

\bibitem{disconnection}
De Ribaupierre, S., Delalande, O., \textit{Hemispherotomy and other disconnective techniques}, Neurosurg. Focus 25: E14 (2008);
Delalande, O., Fohlen, M., \textit{Disconnecting surgical treatment of hypothalamic hamartoma in children and adults with refractory epilepsy and proposal of a new classification},
Neurol. Med. Chir. 43, 61- 68 (2003); Sugano, H., Nakanishi, H., Nakajima, M., Higo, T., Iimura, Y., Tanaka, K., Hosozawa, M., Niijima, S., Arai, H., \textit{Posterior quadrant disconnection surgery 
for Sturge-Weber syndrome}, Epilepsia 55, 683-689 (2014); Mohamed, A. R., Freeman, J. L., Maixner, W., Bailey, C. A., Wrennall, J. A., Harvey, A. S., \textit{Temporoparietooccipital disconnection in 
children with intractable epilepsy}, J. Neurosurg. Pediatr. 7, 660-670 (2011); Dorfer, C., Czech, T., M\"{u}hlebner-Fahrngruber, A., Mert, A., Gr\"{o}ppel, G., Novak, K., Dressler, A., Reiter-Fink, E., 
Traub-Weidinger, T., Feucht, M., \textit{Disconnective surgery in posterior quadrantic epilepsy: experience in a consecutive series of 10 patients}, Neurosurg. Focus 34: E10 (2013);
Hufnagel, A., Zentner, J., Fernandez, G., Wolf, H. K., Schramm, J., Elger, C. E., \textit{Multiple subpial transection for control of epileptic seizures: effectiveness and safety}, Epilepsia 38(6):678-88 (1997);
Ntsambi-Eba, G., Vaz, G., Docquier, M. A., van Rijckevorsel, K., Raftopoulos, C., \textit{Patients with refractory epilepsy treated using a modified multiple subpial transection technique}, Neurosurgery 
72(6):890-7 (2013).

\bibitem{Gonzalez-Martinez2016}
Ross, L., Naduvil, A., Mullin, J., Bulacio, J., Jirsa, V., Chauvel, P. and Gonzalez-Martinez, J., \textit{Modulating Large-Scale Epileptic Networks by SEEG Guided-Laser Ablations}, 
J. Neurosurg. 126(4), A1387-A1387, (2017). 

\bibitem{Jirsa2014}
Jirsa, V. K., Stacey, W. C., Quilichini, P. P., Ivanov, A. I., Bernard C. \textit{On the nature of seizure dynamics}, Brain 137:2210 -2230 (2014).

\bibitem{Proix2014} 
Proix, T., Bartolomei, F., Chauvel, P., Bernard, C., Jirsa, V. K. \textit{Permittivity Coupling across Brain
Regions Determines Seizure Recruitment in Partial Epilepsy}, J. Neurosci. 34, 15009-15021 (2014).

\bibitem{Spiegler2016}
Spiegler, A., Hansen, E.C., Bernard, C., McIntosh, A.R. and Jirsa, V.K., \textit{Selective activation of resting-state networks following focal stimulation in a connectome-based network model 
of the human brain}, eNeuro 3(5), 68 (2016).

\bibitem{Gregoriou2009}
Gregoriou, G. G., Gotts, S. J., Zhou, H., Desimone, R., \textit{High-frequency, long-range coupling between prefrontal and visual cortex during attention}, Science, 324(5931), 1207-1210 (2009).

\bibitem{Grothe2012}
Grothe, I., Neitzel, S. D., Mandon, S., Kreiter, A. K., \textit{Switching neuronal inputs by differential modulations of gamma-band phase-coherence}, J. Neurosci., 32(46), 16172-16180 (2012).

\bibitem{abnormalPZ}
Bower, M.R., Stead, M., Meyer, F.B., Marsh, W.R., and Worrell, G.A. Epilepsia 53, 807-816 (2012); Nair, D.R., Mohamed, A., Burgess, R., and L\"{u}ders, H. Epileptic Disord. 6, 77-83 (2004).

\bibitem{bassett2016}
Khambhati, A.N., Davis, K.A., Lucas, T.H., Litt, B. and Bassett, D.S. \textit{Virtual cortical resection reveals push-pull network control preceding seizure evolution}, 
Neuron, 91(5), pp.1170-1182 (2016).

\bibitem{virtual_resection}
Alstott, J., Breakspear, M., Hagmann, P., Cammoun, L., and Sporns, O. \textit{Modeling the impact of lesions in the human brain}, PLoS Comput. Biol. 5, e1000408 (2009); 
Honey, C.J., and Sporns, O. \textit{Dynamical consequences of lesions in cortical networks}, Hum. Brain Mapp. 29, 802-809 (2008);
V\'{a}\v{s}a, F., Shanahan, M., Hellyer, P.J., Scott, G., Cabral, J. and Leech R. \textit{Effects of lesions on synchrony and metastability in cortical networks}, NeuroImage, 118, pp.456-467 (2015);
Aerts, H., Fias, W., Caeyenberghs, K. and Marinazzo, D. \textit{Brain networks under attack: robustness properties and the impact of lesions}, Brain, p.aww194 (2016).


\bibitem{Burns2014}
Burns, S.P., Santaniello, S., Yaffe, R.B., Jouny, C.C., Crone, N.E., Bergey, G.K., Anderson, W.S. and Sarma, S.V. \textit{Network dynamics of the brain and influence of the epileptic seizure onset zone},
Proceedings of the National Academy of Sciences, 111(49), pp.E5321-E5330 (2014).

\bibitem{Taylor2014}
Taylor, P.N., Wang, Y., Goodfellow, M., Dauwels, J., Moeller, F., Stephani, U. and Baier, G. \textit{A computational study of stimulus driven epileptic seizure abatement}, PLOS one, 9(12), p.e114316 (2014).

\bibitem{Sinha2017}
Sinha, N., Dauwels, J., Kaiser, M., Cash, S.S., Brandon Westover, M., Wang, Y. and Taylor, P.N. \textit{Predicting neurosurgical outcomes in focal epilepsy patients using computational modelling}, 
Brain, 140(2), pp.319-332 (2016).

\bibitem{abnormalEZ}
Jiruska, P., de Curtis, M., Jefferys, J.G.R., Schevon, C.A., Schiff, S.J., and Schindler, K. J. \textit{Synchronization and desynchronization in epilepsy: controversies and hypotheses}, Physiol. 591, 787-797 (2013); 
Kramer, M.A., and Cash, S.S. \textit{Epilepsy as a disorder of cortical network organization}, Neuroscientist 18, 360-372 (2012); 
Kramer, M.A., Eden, U.T., Kolaczyk, E.D., Zepeda, R., Eskandar, E.N., and Cash, S.S. \textit{Coalescence and fragmentation of cortical networks during focal seizures}, J. Neurosci. 30, 10076-10085 (2010).




\end{thebibliography}
\end{document}